# The ArTéMiS wide-field submillimeter camera: preliminary on-sky performances at 350 microns.

Vincent Revéret[*,a], Philippe André[a], Jean Le Pennec[a], Michel Talvard[a], Patrick Agnèse[b], Agnès Arnaud[b], Laurent Clerc[c], Carlos de Breuck[d], Jean-Charles Cigna[b], Cyrille Delisle[a], Eric Doumayrou[a], Lionel Duband[c], Didier Dubreuil[a], Luc Dumaye[a], Eric Ercolani[c], Pascal Gallais[a], Elodie Groult[a], Thierry Jourdan[c], Bernadette Leriche[e], Bruno Maffei[f], Michel Lortholary[a], Jérôme Martignac[a], Wilfried Rabaud[b], Johan Relland[g], Louis Rodriguez[a], Aurélie Vandeneynde[b], François Visticot[a]

[a]AIM Paris Saclay, UMR CEA/CNRS/UP7, CEA/Irfu/SAp, Bât 709, Orme des Merisiers, 91191 Gif-sur-Yvette, France; [b]CEA-LETI/SLIR 17 rue des Martyrs, 38054 Grenoble, France; [c]CEA/DSM/SBT 17 rue des Martyrs, 38054 Grenoble, France; [d]ESO Garching Karl-Schwarzschild-Str. 2, 85748 Garching bei München, Germany; [e]IAS, Université Paris Sud, 91405 Orsay, France; [f]JBCA, School of Physics and Astronomy, The University of Manchester, Alan Turing Building, Oxford road, Manchester M13 9PL, UK; [g]CEA/Irfu/SIS, 91191 Gif-sur-Yvette, France

**ABSTRACT**

ArTeMiS is a wide-field submillimeter camera operating at three wavelengths simultaneously (200, 350 and 450 µm). A preliminary version of the instrument equipped with the 350 µm focal plane, has been successfully installed and tested on APEX telescope in Chile during the 2013 and 2014 austral winters. This instrument is developed by CEA (Saclay and Grenoble, France), IAS (France) and University of Manchester (UK) in collaboration with ESO.

We introduce the mechanical and optical design, as well as the cryogenics and electronics of the ArTéMiS camera. ArTeMiS detectors consist in Si:P:B bolometers arranged in 16×18 sub-arrays operating at 300 mK. These detectors are similar to the ones developed for the Herschel PACS photometer but they are adapted to the high optical load encountered at APEX site. Ultimately, ArTeMiS will contain 4 sub-arrays at 200 µm and 2×8 sub-arrays at 350 and 450 µm. We show preliminary lab measurements like the responsivity of the instrument to hot and cold loads illumination and NEP calculation.

Details on the on-sky commissioning runs made in 2013 and 2014 at APEX are shown. We used planets (Mars, Saturn, Uranus) to determine the flat-field and to get the flux calibration. A pointing model was established in the first days of the runs. The average relative pointing accuracy is 3 arcsec. The beam at 350 µm has been estimated to be 8.5 arcsec, which is in good agreement with the beam of the 12 m APEX dish. Several observing modes have been tested, like "On-The-Fly" for beam-maps or large maps, spirals or raster of spirals for compact sources. With this preliminary version of ArTeMiS, we concluded that the mapping speed is already more than 5 times better than the previous 350 µm instrument at APEX. The median NEFD at 350 µm is 600 mJy.s$^{1/2}$, with best values at 300 mJy.s$^{1/2}$. The complete instrument with 5760 pixels and optimized settings will be installed during the first half of 2015.

## 1. INTRODUCTION

Submillimeter astronomy is usually defined within the spectral range ~ 0.2 – 1 mm, and is associated with the observation of the cold Universe. The evolution of star formation rate through the Universe can be studied by mapping cold (~10 K) gas and dust regions in the submillimeter. The continuum emission from these regions is usually optically thin, giving access to the central parts, where recently, gas filaments have been observed by Herschel[1].

---

[*] Contact : vincent.reveret@cea.fr

The densest filaments contain pre-stellar cores that are the seeds of future stars. The scenario of star formation is getting now more and more known, but still requires follow-up observations to constrain the dynamics of the filamentary structures. Large surveys, with moderate spatial resolution are necessary to map molecular clouds at different distances from Earth, in order to verify if the star-formation scenario is universal. ALMA is the perfect instrument to perform such analysis on small angular scales, but it is quite limited to conduct large fields mapping. This is the reason why submillimeter continuum cameras with high mapping speeds are needed, in perfect complementarity to a submillimeter interferometer such as ALMA.

ArTeMiS is a submillimeter camera specifically designed to interface with the 12 meters aperture APEX[8] submillimeter telescope located at the Llano de Chajnantor in the Atacama desert in Chile. The site is at an altitude of 5100 meters, where the atmospheric pressure is about half of that at sea level, and it experiences some of the lowest precipitable water vapor (pwv) conditions available anywhere in the world. This makes it suitable for performing ground-based submillimeter observations through atmospheric windows, even possibly within the challenging 200 µm band. The instrument follows up on the precursor instrument P-ArTeMiS[2], which has been already tested on the APEX telescope in 2007 and 2009 although it performed astronomical observations with only one spectral band at 450 µm.

The detectors are planar bare arrays of silicon bolometers, which do not use feed horns to concentrate the submillimeter signals. ArTeMiS includes 20 sub-arrays of 16×18 pixels operating simultaneously (4 arrays at 200 µm, 8 arrays at 350 µm and 8 arrays at 450 µm, 5760 pixels in total). The field of view for the 350 and 450 µm focal planes is 4.7×2.3 arcmin$^2$, and 2.3×2.3 arcmin$^2$ for the 200 µm channel. The following section describes the system: optics, filtering, cryogenics and electronics. The next sections describe the detectors and the laboratory tests, as well as the preliminary on-sky measurements at APEX. We performed 3 commissioning and scientific runs between 2013 and 2014, with only the 350 µm focal plane installed. The last section presents the future plans for ArTéMiS (installation of the remaining detectors, improvement of the sensitivity).

ArTéMiS is a collaborative project, led by the astrophysics division of CEA Saclay, France, with important contribution from CEA Grenoble (LETI and SBT), Manchester University in UK, Institut d'Astrophysique Spatiale (IAS) and Institut d'Astrophysique de Paris (IAP) both in France. The project received a lot of support from European Southern Observatory (ESO), APEX staff and Onsala Space Observatory in Sweden. ArTéMiS was partially funded by a French Grant ANR.

## 2. INSTRUMENT DESIGN

ArTéMiS is installed in the Cassegrain cabin of the APEX telescope. It is mounted on the roof of the cabin and has a non-conventional shape in order to fit the small allocated space.

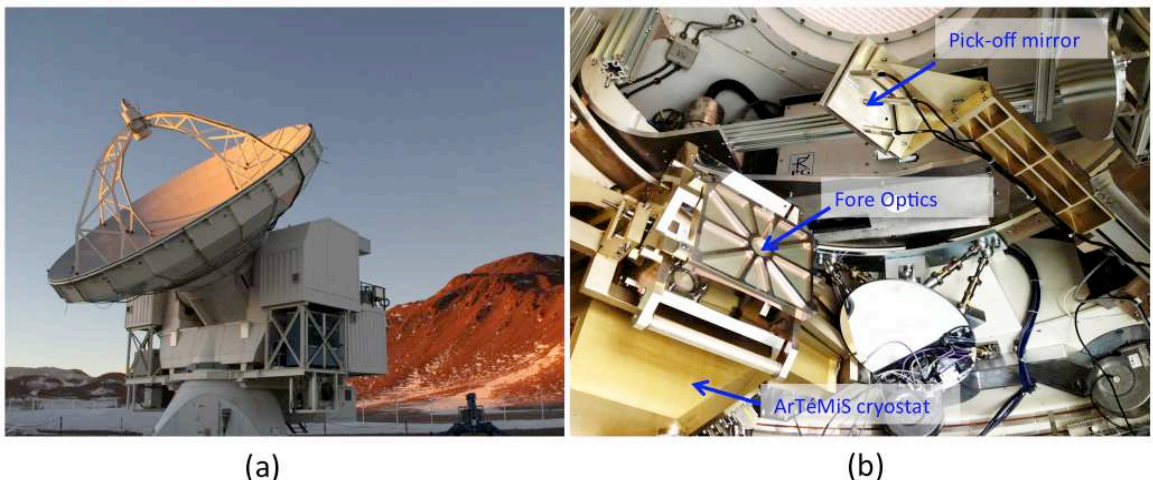

Figure 1: (a) The APEX telescope, a 12m single dish in the Atacama desert, 5105m above sea level. APEX is operated by ESO, and funded by MPIfR, ESO, OSO and Chile. (b) ArTéMiS inside the Cassegrain cabin of APEX. The pick-off mirror is mounted at the end of a moving arm and sends the beam to the ArTéMiS Fore-Optics.

## 2.1 Optics and Filtering

The core of ArTeMiS is the bolometer arrays. Because ArTeMiS doesn't use feed horns, the theory of Gaussian beam modes and Gaussian beam propagation was not employed, as it is the case for main sub millimeter systems, to optimize the optical design. Our approach was rather driven by our experience in mid-infrared instrumentation[4]. The result of this approach is the use of high-quality reflective optics all along the optical path.

The instrument is made of:

- A folding mirror (pick-off mirror) linked to the Cassegrain cabin. This mirror is adjustable in tilt, through 3 mechanisms, to match both telescope and ArTeMiS optical axis.

- A Fore Optics working at room temperature. It is composed of one additional folding mirror and two toroidal mirrors. The Fore Optics is considered as a rigid subset adjustable with respect to the ArTeMiS cryostat with 6 degrees of freedom.

- A cryostat including the Cold Optics for the 3 beams and the filters. Inside the cryostat two dichroic plates are required to separate the bandwidths. Both optical paths at 200 µm and (350 µm and 450 µm) are split up by the first dichroic plate working at 60 K. The 200 µm beam is transmitted while the (350 µm and 450 µm) beam is reflected. Then the 350 µm and 450 µm beams are split up by the second dichroic plate working at 4K. The 450 µm beam is transmitted while the 350 µm is reflected (see figure 2).

To allow ArTeMiS self-alignment and alignment with respect to the telescope optical axis, dedicated tools, a lens alignment at the 350 µm and 450 µm cold stop and a diaphragm at the intermediate focal plane, both using mechanisms, are implemented in the cryostat. More details about the optical design are found in Dubreuil[5] et al, in these proceedings.

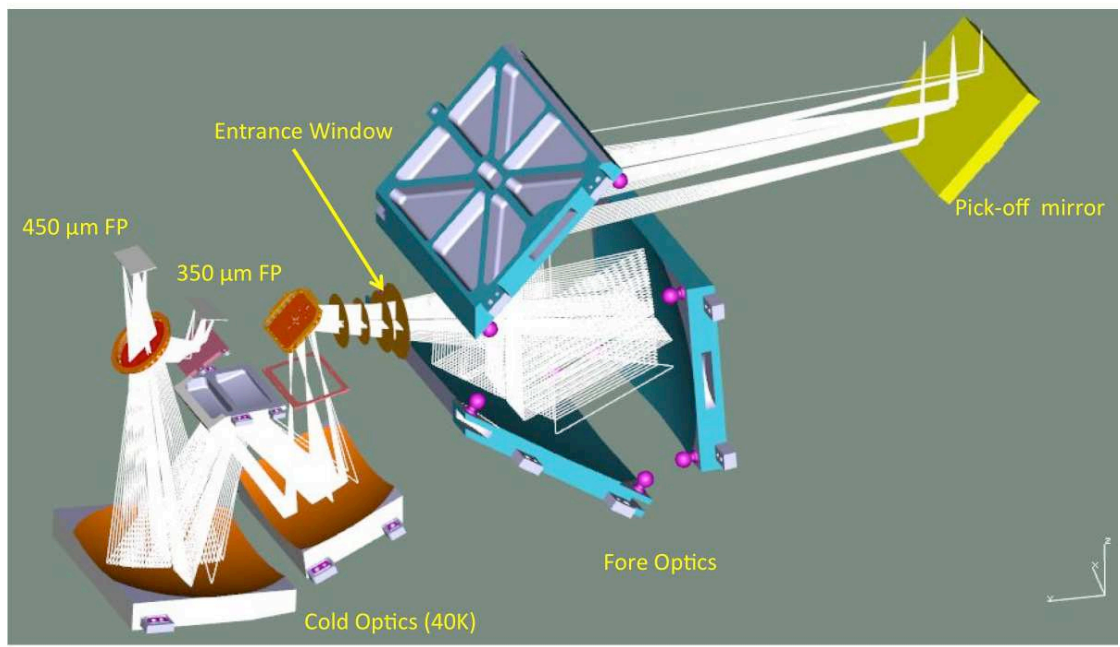

Figure 2: 3D view of the ArTéMiS all-reflective optics. The Fore Optics is at room temperature, and folds the beam to enter the cryostat. The 200 µm focal plane is not represented in this view.

The optical elements divide into four main function categories, which are cryostat pressure window, thermal rejection, channel separation and channel definition. The local mounting temperature of each element can be stated but that may not be the actual temperature of the element due to balancing of radiant heat fluxes. All the elements are relatively large diameter, so lateral thermal conduction is not expected to be significant as a mode of heat loss.

Possibly due to design limitations, no single type of filter can be configured to fit all the roles required in such a camera system. Without either of these developments, improvements can still be made by developing automatic code, which designs the whole system rather than just the individual elements. Only then will the combination of all the parts give a fully optimal system.

Even though a further iteration on optimization of filtering will be performed for future observation runs, the overall performance of the instrument during the first run has shown to be very competitive. A description of the filtering scheme for ArTéMiS is found in Haynes et al[9].

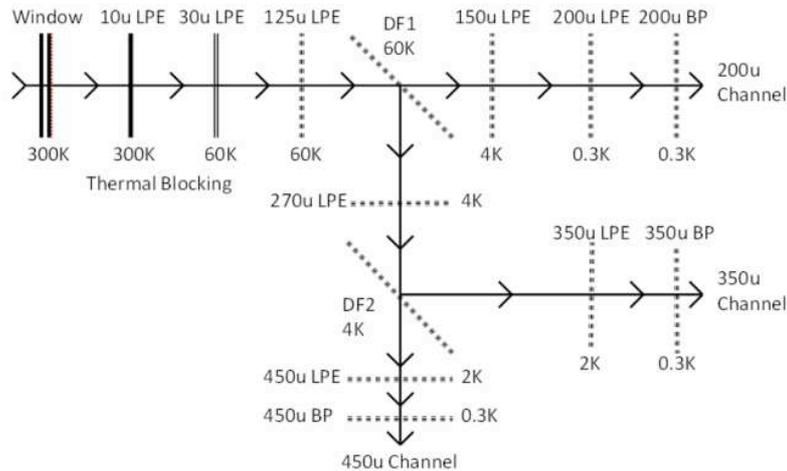

Figure 3: Schematic of the ArTeMiS filtering layout, displaying the local temperature at each element. LPE stands for Low Pass Edge and BP stands for Band Pass.

## 2.2 Cryogenics

The ArTéMiS cryostat has been developed at the SBT, the low temperature laboratory of the INAC institute in CEA Grenoble, France. To cope with the reduced and curved space allocated within the telescope, the cryostat exhibits an unusual compact geometry. The main consequence is the complex shapes of the thermal shields that are obtained using an original aluminum manufacturing process "the solder bath salt". The thermal and mechanical features of such aluminum are measured in laboratory through many tests at ambient and low temperatures and are consistent with the needs. The inner structure of the cryostat is a 40 K plate which acts as an optical bench and is bound to the external vessel through two hexapods, one fixed and the other one mobile thanks to a ball bearing. Once the cryostat is cold, this characteristic enables all the different elements to be aligned with the optical axis.

The cryogenic chain is built around a pulse tube cooler (40 K and 4 K) coupled to a double stage helium sorption cooler (300 mK). The temperature of the pulse tube varies depending on the telescope elevation (15° to 90°) but the relative thermal insulation of the sorption cooler evaporator stage gives a stable 0.3 K stage. All the vacuum and cooling processes can be controlled through a system based on a *Program Logic Controller* (PLC) either in automatic or in manual mode. If the link between the telescope and the control room is lost or broken, the PLC reliably takes over and maintains the cryostat in a safe condition. The PLC is connected with several devices through several protocols (Profibus DP, USB, etc). It is linked to a *Supervisory Control and Data Acquisition* (a SCADA named Muscade®) which is integrated in a fan-less PC. The whole configuration of this PC is written in a flash memory card. More details on the cryogenic system for ArTéMiS can be found in reference Ercolani et al[7].

The typical time between two recycling procedures at the telescope is 48 hours. The maximum hold time at 260mK under operation at APEX with the entire 350 μm focal plane is 84 hours. No extra noise due to the pulse-tube has been observed on the bolometer signal.

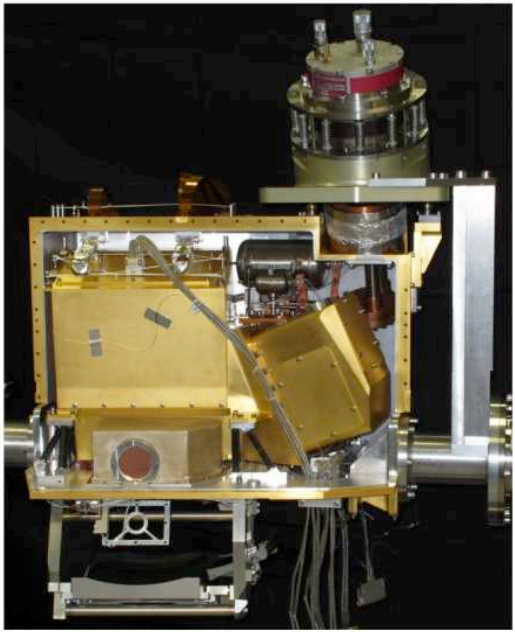 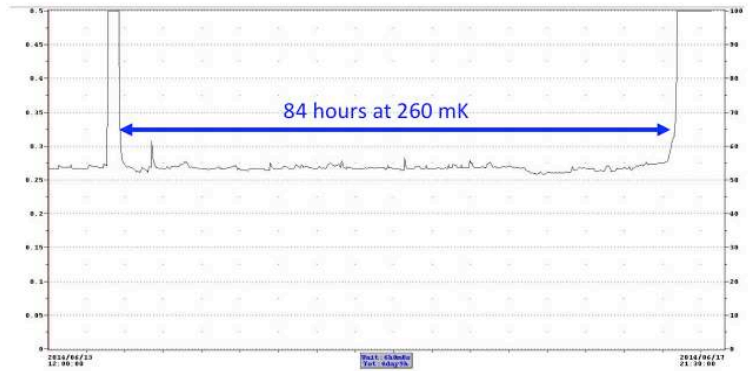

Figure 4: (a) Inside view of the ArTéMiS cryostat. The pulse tube is visible on the top, as well as the different shields for optics and detectors. (b) The maximum hold time at APEX at 260 mK, in June 2014.

## 2.3 Readout electronics

During ArTeMiS observations at the APEX telescope the arrays have to be read out simultaneously at 40Hz. The readout electronics consists of cryogenic buffers at 4K (NABU), based on CMOS technology, and of warm electronic acquisition systems called BOLERO. The bolometric signal given by each pixel has to be amplified, sampled, converted, time stamped and formatted in data packets by the BOLERO electronics. The time stamping is obtained by the decoding of an IRIG-B signal given by APEX and is a key element to ensure the synchronization of the data with the telescope. Specifically developed for ArTeMiS, BOLERO is an assembly of analogue and digital FPGA boards connected directly on the top of the cryostat. Two detectors arrays (18×16 pixels), one NABU and one BOLERO interconnected by ribbon cables constitute the unit of the electronic architecture of ArTeMiS. In total, the 20 detectors for the three focal planes are read by 10 BOLEROs.

The software is working on a Linux operating system. It runs on 2 back-end computers (called BEAR) which are small and robust PCs with solid state disks. They gather the 10 BOLEROs data fluxes, and reconstruct the focal planes images. When the telescope scans the sky, the acquisitions are triggered thanks to a specific network protocol. This interface with APEX enables to synchronize the acquisition with the observations on sky: the time stamped data packets are sent during the scans to the APEX software that builds the observation FITS files. A graphical user interface enables the setting of the camera and the real time display of the focal plane images, which is essential in laboratory and commissioning phases. The software is a set of C++, Labview and Python, the qualities of which are respectively used for rapidity, powerful graphic interfacing and scripting. The commands to the camera can be sequenced in Python scripts. The complete description of the readout system is available in Doumayrou et al[3].

## 3. DETECTORS

ArTeMiS detectors consist in Si:P:B bolometers arranged in 16×18 pixels sub-arrays operating at 300 mK or a bit below. These detectors are similar to the ones developed for the Herschel PACS photometer (60-210 µm) but they are adapted to longer wavelength and to the high optical load encountered at APEX site. The submillimeter radiation is efficiently absorbed by each bolometer thanks to a superconducting TiN layer deposited on a silicon grid above a backshort[10]. The size of the cavity is defined by indium bumps with a size of 50 μm. The spectral response has been measured in the lab and is above 92% in the range 200 – 550 μm.

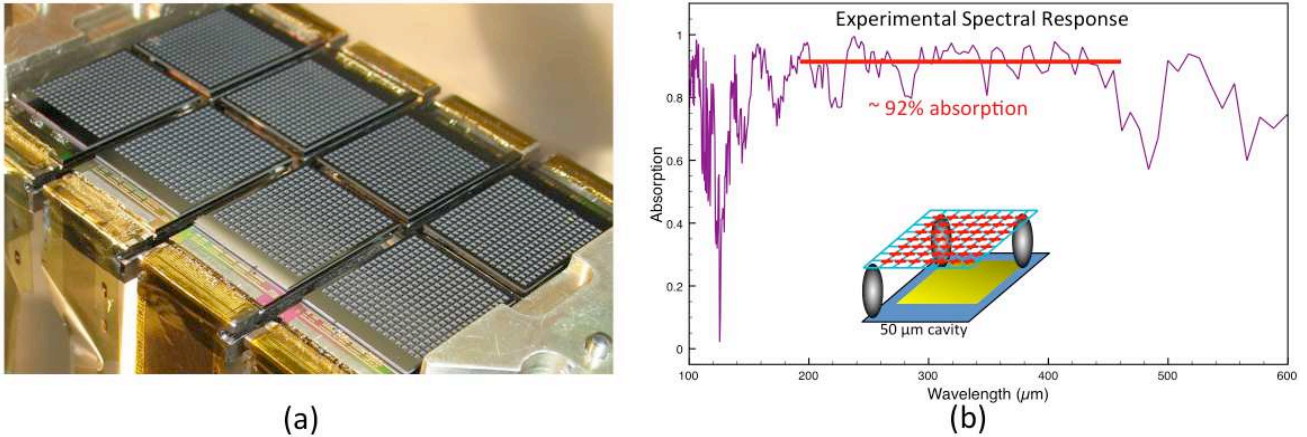

Figure 5: (a) The ArTéMiS 350 μm focal plane used for the 2014 run at APEX. The 300 mK ribbon cables visible on one side of each 16×18 array connect the arrays to the 4K electronic stage (NABU). (b) Spectral response of the ArTéMiS bolometers measured by an FTS in the lab. The average absorption is 92%.

These bolometers[11] have very large impedance (of the order of 100 GΩ at 300 mK) and also very high responsivity (> $2\times10^{10}$ V/W at 300 mK) because they operate in the Efros-Shklovskii variable range hopping regime[6]. To get a reasonable value for the time constant of the detectors (around 80 ms), a cryogenic impedance matching stage is necessary close to the detection stage. We use CMOS transistors at 300 mK, that also perform the multiplexing (16 to 1, time domain multiplexing).

We performed tests prior to the shipment to Chile early 2014, in order to verify the overall behavior of the instrument and to give first estimations of the sensitivity. On the inset of figure 6, we show the amplitude of the signal (in Volt) for each pixel, corresponding to the difference between a hot (room temperature Eccosorb) and cold (80 K Eccosorb) illumination. We see that the focal plane has a yield equal to ~ 90%. We can distinguish between two types of bad pixels: individual ones and complete lines (they usually correspond to disconnected cryogenic bonding). The laboratory Noise Equivalent Power (NEP) has been measured with a background power of 25 pW per pixel, which corresponds to medium atmospheric conditions at 350 μm at APEX. The median value at 300 mK is $NEP_{lab} = 4.5\times10^{-16}$ W.Hz$^{-1/2}$, the best pixel has $NEP_{best} = 2.7\times10^{-16}$ W.Hz$^{-1/2}$.

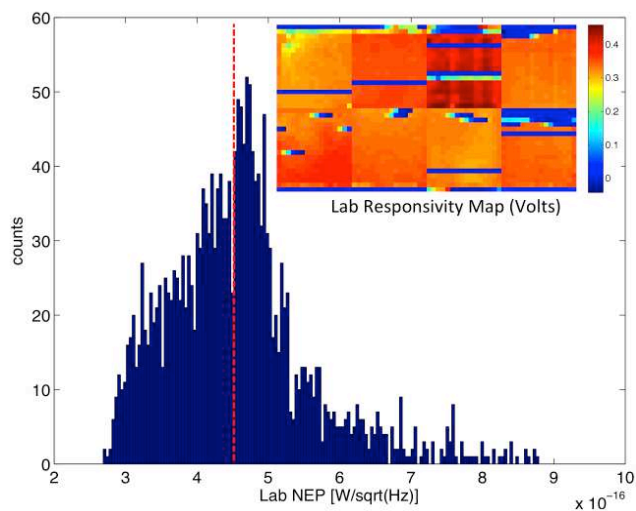

Figure 6: Histogram of laboratory NEPs for the 350 μm focal plane. The inset shows the map of response (amplitude of the output signal) to a Hot / Cold illumination at the entrance of the camera. This gives a yield of 90%.

## 4. ON-SKY OPERATIONS

ArTéMiS has first been installed on APEX in July 2013 with 4 arrays at 350 µm only. In May 2014, 4 more arrays at 350 µm have been mounted and the commissioning started in the beginning of June. With the help from APEX staff, the installation of the instrument in the Cassegrain cabin, the setting of the cryogenic equipment, the optical alignment and the electronics installation took only 4 days despite the bad weather (snow, strong wind and very low temperature). The integration of ArTéMiS data stream into the APEX network was straightforward, as well as the integration in the control system. We developed specific Python scripts for specific maps, or combinations of particular scans types and offsets. The following parts describe the preliminary results obtained by ArTéMiS 350 µm on APEX.

### 4.1 Observing Modes

ArTéMiS can be used with all the major observing modes available at APEX. The figure 7a shows for example the on-sky trajectory (in horizontal coordinates) of a "spiral map". This kind of map is used to correct for pointing offsets on strong sources (planets, HII regions). For larger fields, we have successfully tested "rasters of spirals", or the classical "On-The-Fly" mode. To characterize the atmospheric opacity, we performed "skydips" like it is the case on other (sub)millimeter telescopes.

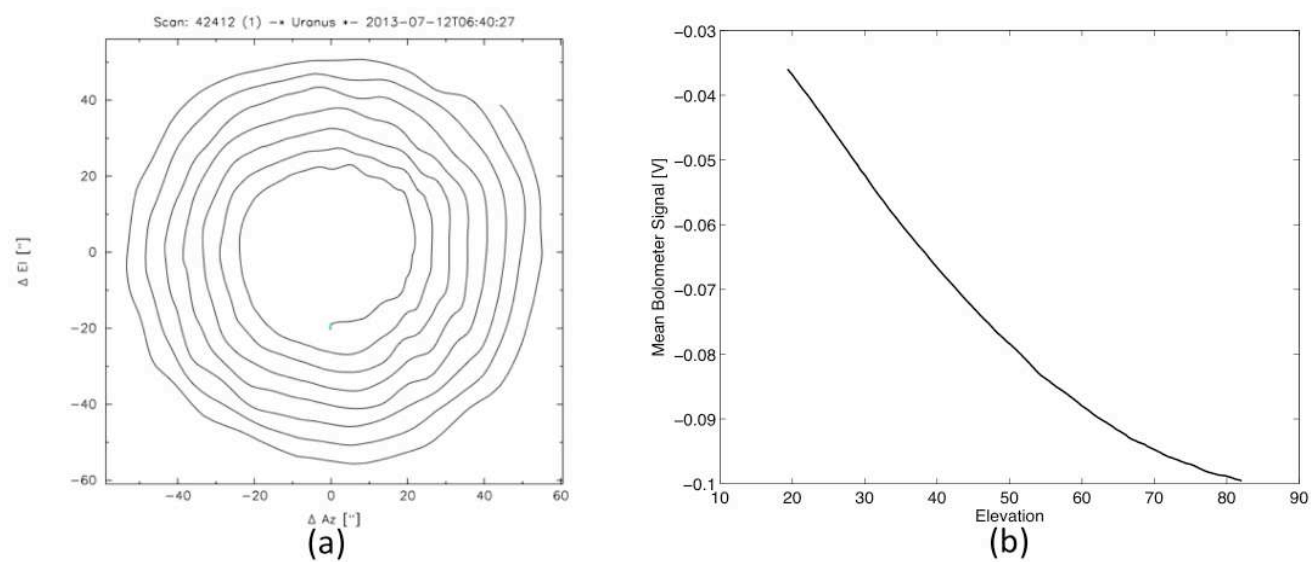

Figure 7: (a) APEX trajectory for a spiral-map in horizontal coordinates, used by ArTéMiS for on-source pointing. (b) Typical continuous signal for one bolometer when APEX is doing a "skydip".

### 4.2 Pointing – Focusing - Calibration

During this run, the measured PFOV was (3.85 ± 0.05) arcsec, in perfect agreement with the nominal value of 3.87 arcsec. The FWHM of the 350 µm image of Mars, which apparent diameter was 4.2 arcsec, was around 8.5 arcsec. It corresponds to a mean value of the PSF, for a 'on-the–fly' mode and a field of view 210 arcsec × 210 arcsec, of 8.1 arcsec. The equivalent nominal value is 6.5 arcsec. The gap of 1.6 arcsec could be explained by other increasing effects as the real APEX telescope PSF, its absolute pointing accuracy of 0.6 arcsec, a residual defocus or lag effect. A pointing model was implemented after the first days of observations and gave very good results (3 arcsec rms precision) all along the observing runs.

We used the APEX procedures to correct the focus during an observing project. The secondary mirror can be moved in x, y, and z positions. A "z" focus adjustment is made every 3 hours, with typical correction of ±0.1 mm. We noted strong dependence of the Sun illumination, as well as an elevation dependence for the "y" and "z" focus correction. This last effect will be investigated much further in the next commissioning runs.

The calibration was made from maps on Mars, Uranus and Neptune. The process is still on going as we continue to observe secondary calibrators.

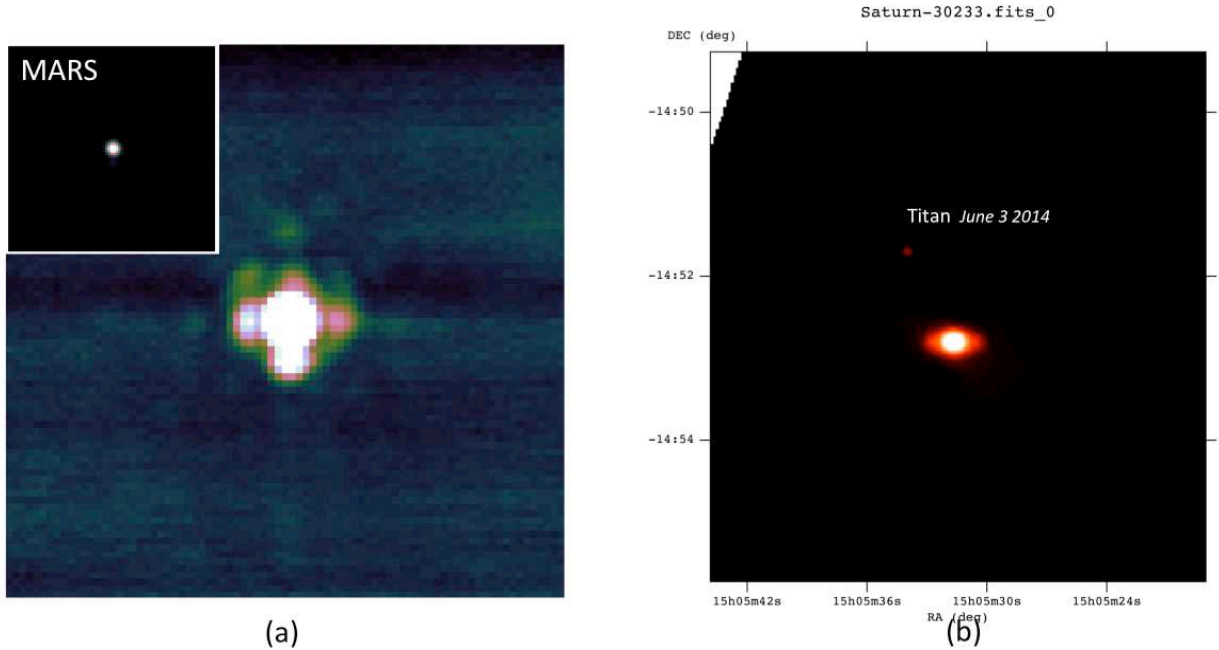

Figure 8: (a) Mars beam-map, with full scale in the inset, and reduced scale in the large image to highlight the error beam. (b) Saturn and its rings, and Titan.

### 4.3 Skynoise

The sky noise at 350 μm is clearly seen on the full focal plane and is highly correlated between the pixels, as it can be observed on figure 9. The data reduction packages like BoA[12], or IDL scripts, can easily remove the correlated part of the signal. The "cleaned" signals clearly have a less important 1/f noise contribution. We also note that there is no evidence of noise (peak) induced by the pulse-tube running frequency around 1 Hz.

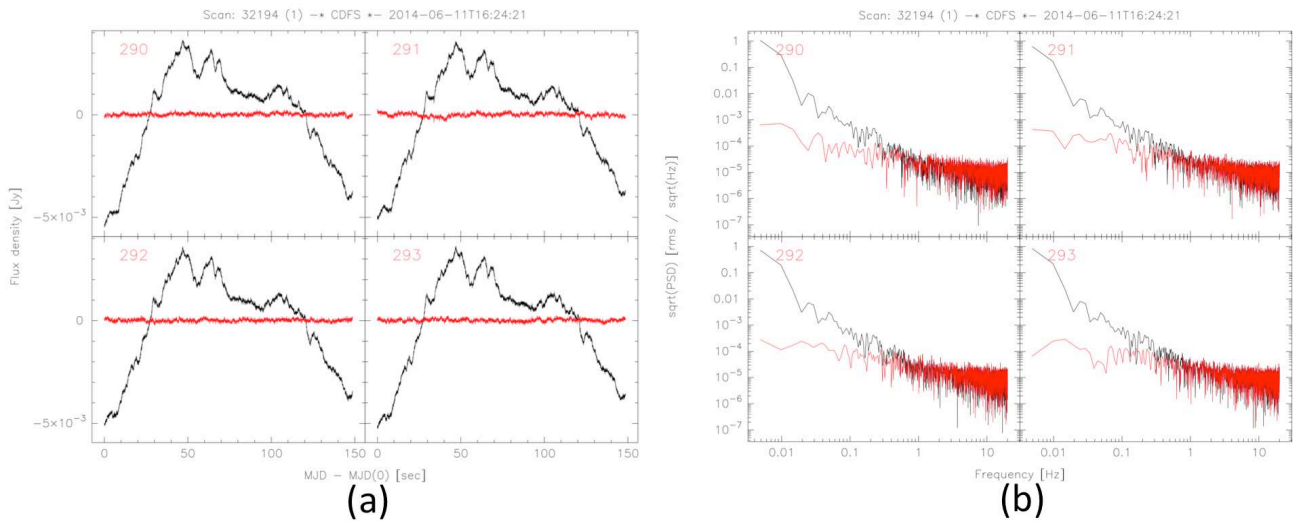

Figure 9: (a) Timeline of 150 s for 4 pixels. The black curves correspond to raw signals. The large fluctuations correspond mainly to sky noise. Signals in red are obtained after the subtraction of the correlated noise. (b) Corresponding noise spectrum for the 4 previous pixels. The 1/f noise has been dramatically removed after decorrelation.

Some work is going on in order to improve the noise reduction, in particular the possible removal of the correlated noise for pixels sharing the same electronic components (for example the 16 pixels from one line that share the same MOSFET transistor, or pairs of sub-arrays that share a single NABU unit).

### 4.4 First results

ArTéMiS is currently in commissioning stage and is not fully operational (one band is present, out of three). But it has started several scientific projects at 350 μm, through ESO and Onsala Space Observatory scientific periods. The two images below show some of these first results. Several star forming regions (extended molecular clouds or smaller regions around dense cores) have been observed, as well as infrared dark clouds, point sources (HII regions mainly) or regions around the galactic center. Observations of deep fields are scheduled for the next runs later in 2014 and will be perfect opportunities to test how the noise evolves with integration time.

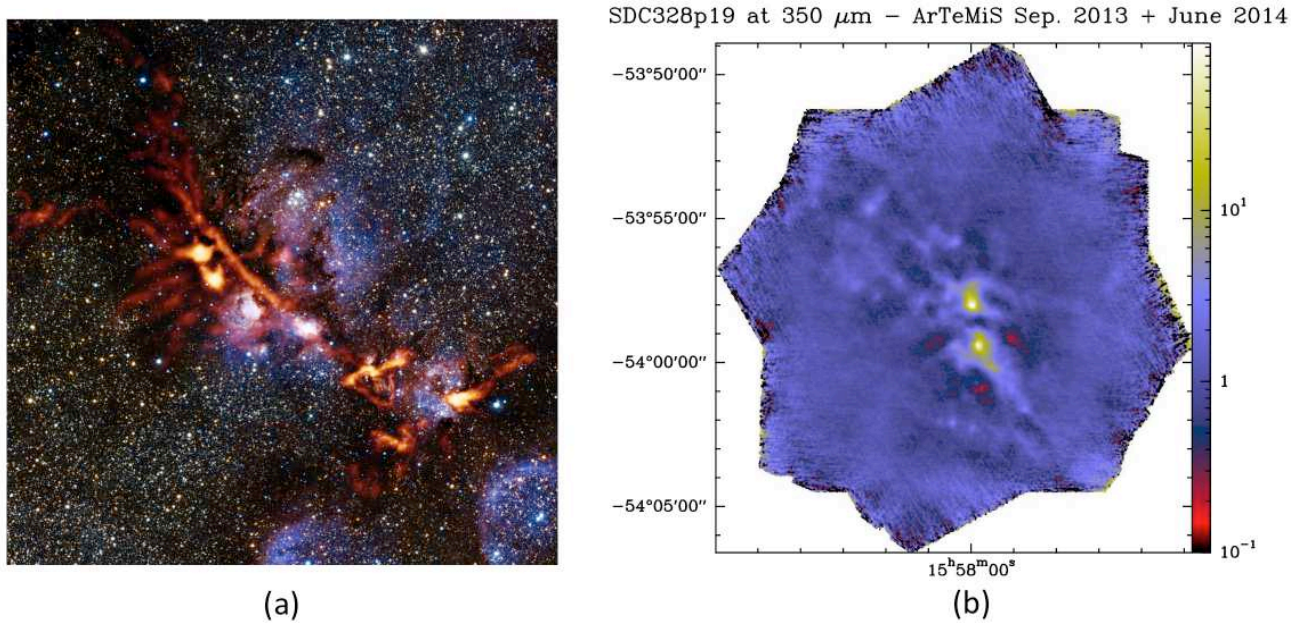

Figure 10 : (a) NGC6334 (the Cat's Paw Nebula) is a star forming region in the Scorpion. The background image comes from VISTA observations, ArTéMiS data at 350 μm are in orange. The filaments are clearly visible around the most active areas, that correspond to pre-stellar cores (PI: P. André CEA, Vista data : J. Emerson). (b) An image of the Spitzer Dark Cloud SDC328, that shows more spatial details than the Herschel view of the same region (PI : N. Peretto, Cardiff University).

### 4.5 Sensitivity

We have determined the sensitivity during the 2014 run, by doing beam maps on Mars and Uranus, under good atmospheric conditions (pwv below 0.3 mm). The median value for the Noise Equivalent Flux Density is NEFD = 600 mJy.s$^{1/2}$, with best values at 300 mJy.s$^{1/2}$. The number of working pixels at APEX is 1650 because one array was not functional, probably because of a connector problem, at the focal plane level.

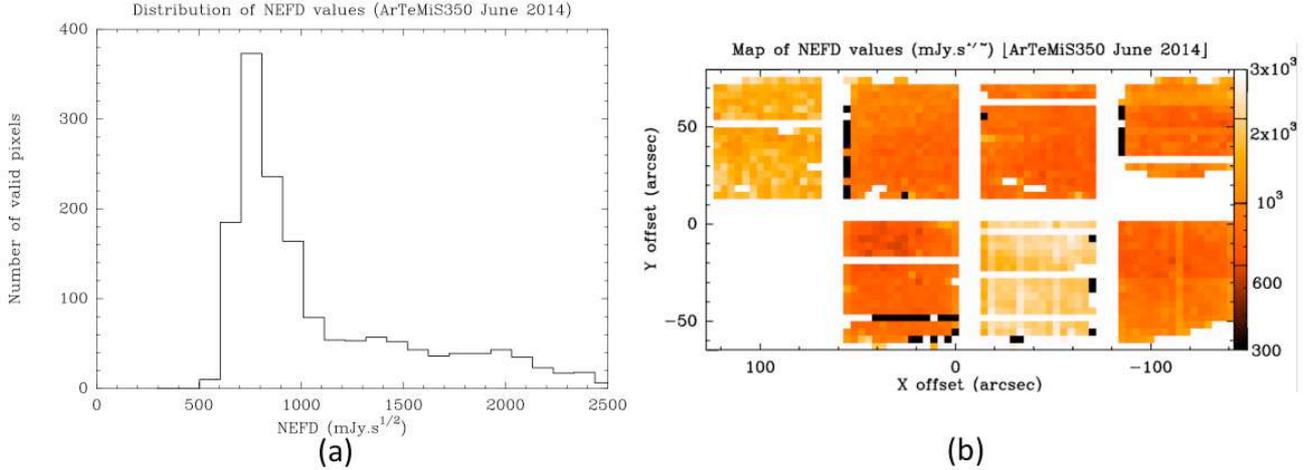

Figure 11: (a) Distribution of the pixels NEFD (mJy.s$^{1/2}$). (b) Spatial distribution of the NEFD. One array is not functional, and 2 others are clearly below the others.

Moreover, 2 arrays have lower than average performances. This problem will be investigated when ArTéMiS will come back to the lab. By comparing with the previous 350 μm instrument at APEX, Saboca[13], the improvement in mapping speed for extended sources is a factor 5 in favor of ArTéMiS with this 2014 version. With a complete version (better 350 μm arrays and complete 200 and 450 μm focal planes), ArTéMiS can be up to 16 times faster than Saboca.

## 5. CONCLUSION AND FUTURE PLANS

ArTéMiS has shown excellent results after the 2013 and 2014 runs at APEX, even with an incomplete instrument. The installation phase, alignment and first on-sky tests went relatively fast and were successful in term of scientific observations. The instrument will be shipped back to CEA, France, in October 2014 to install the remaining arrays. We plan to optimize the optical filtering in order to reduce the optical load on the pixels. If successful, this change will improve the responsivity. We clearly see in the lab that going from a background power per pixel of 35 pW (estimated value for pwv = 0.3mm at APEX, with current filtering) to 25 pW (estimated value for the same pwv, with optimized filtering), the responsivity (in V/W) is improved by 35%. The full instrument will be installed at the beginning of 2015.

| | |
|---|---|
| **Number of operational pixels** | 1650 |
| **Spatial Resolution @ 350 μm** | 8'' |
| **FOV** | 4,7 × 2,3 arcmin$^2$ |
| **Median NEFD** | 600 mJy.s$^{1/2}$ |
| **Best NEFD** | ~ 300 mJy.s$^{1/2}$ |
| **Mapping speed (relatively to Saboca)** | ×5 (×16 if extrapolated to full instrument) |

Table 1: Summary of ArTéMiS performances at 350 μm, for the 2014 runs at APEX.

## 6. ACKNOWLEDGEMENTS

We would like to thank the local staff at APEX that continuously helped us during the 2 installation phases, despite the sometimes-bad weather conditions. We also thank people at ESO Garching for their support and interest and help in data reduction during the commissioning. ArTéMiS received a grant from the French *Agence Nationale de la Recherche*, ANR, ref. ANR-05-BLAN-0215.